\documentclass[apj]{emulateapj}
\usepackage{graphicx, longtable, latexsym, amssymb, lscape, natbib, color}
\begin{document}

\newcommand{\scprime}{{\sc prime}}
\newcommand{\Bprime}{{\sc Bprime}}
\newcommand{\Hprime}{{\sc Hprime}}
\newcommand{\HIprime}{{\sc Gasprime}}
\newcommand{\field}{{\sc Field}}
\newcommand{\cluster}{{\sc Virgo}}
\newcommand{\sdssnair}{{\sc NA10}}
\newcommand{\FIELD}{{\sc Field+}}
\newcommand{\CLUSTER}{{\sc Cluster+}}

\title{Multi-Wavelength Properties of Barred Galaxies in the Local
  Universe: Environment and evolution across the Hubble
  sequence}

\author{Lea Giordano\altaffilmark{1}, Kim-Vy H.
  Tran\altaffilmark{1,2}, Ben Moore\altaffilmark{1} \& Am\'{e}lie
  Saintonge\altaffilmark{3,4}} \email{giordano@physik.uzh.ch}

\altaffiltext{1}{Institute for Theoretical Physics, University of
  Z\"{u}rich, Switzerland} 

\altaffiltext{2}{George P. and Cynthia W. Mitchell Institute for
Fundamental Physics and Astronomy, Department of Physics \& Astronomy,
Texas A\&M University, College Station, TX 77843}

\altaffiltext{3}{Max Planck Institute for Astrophysics, Garching,
  Germany}

\altaffiltext{4}{Max-Planck-Institut f\"{u}r Extraterrestrische
  Physik, Garching, Germany}

\date{\today}

\begin{abstract}

  We investigate possible environmental and morphological trends in
  the $z\sim0$ bar fraction using two carefully selected samples
  representative of a low-density environment (the isolated galaxies
  from the AMIGA sample) and of a dense environment (galaxies in the
  Virgo cluster). Galaxies span a stellar mass range from $10^8$ to
  $10^{12}$M$_{\odot}$ and are visually classified using both
  high-resolution NIR (H-band) imaging and optical \texttt{rgb}
  images. We find that the bar fraction in disk galaxies is
  independent of environment suggesting that bar formation may occur
  prior to the formation of galaxy clusters.  The bar fraction in
  early type spirals ($Sa-Sb$) is $\sim$50\%, which is twice as high
  as the late type spirals ($Sbc-Sm$). The higher bar fraction in
  early type spirals may be due to the fact that a significant
  fraction of their bulges are pseudo-bulges which form via the
  buckling instability of a bar. i.e. a large part of the Hubble
  sequence is due to secular processes which move disc galaxies from
  late to early types.  There is a hint of a higher bar fraction with
  higher stellar masses which may be due to the susceptibility to bar
  instabilities as the baryon fractions increase in halos of larger
  masses.  Overall, the $S0$ population has a lower bar fraction than
  the $Sa-Sb$ galaxies and their barred fraction drops significantly
  with decreasing stellar mass. This supports the notion that $S0s$
  form via the transformation of disk galaxies that enter the cluster
  environment.  The gravitational harassment thickens the stellar
  disks, wiping out spiral patterns and eventually erasing the bar - a
  process that is more effective at lower galaxy masses.

\end{abstract}


\maketitle

\section{Introduction}
\label{sec:intro}

\setcounter{footnote}{0}

Understanding bar-formation and the secular effects that bars have on
the stellar component is becoming central to our understanding of
galaxy formation and evolution.  Once a bar forms it can change the
scale length of the stellar component via scattering/mixing of stars
in the radial direction and they can create an extended stellar
distribution \citep{Roskar2008}.  Bars can move a galaxy between
morphological classes through the secular formation of pseudo-bulges
\citep{Athanassoula2002, Debattista2004} and they can drive gas to the
central black hole fuelling AGN (Shlosman et al. 1989).

Numerical simulations have shown that bars naturally arise from the
secular evolution of discs \citep{Toomre1964, Ostriker1973,
  Fall1980}. Bars can also be triggered by dynamical interactions in
the field \citep{Gerin1990, Barnes1992, RomanoDiaz2008, Dubinski2009}.
It has been shown that close galaxy companions are associated with bar
formation, but primarily for early Hubble-types \citep{Elmegreen1990}.
In galaxy clusters, gravitational encounters (harassment) can drive a
morphological transformation from late type disks to dwarf ellipticals
(dSphs). In this scenario, encounters create a ``naked'' stellar bar
which is subsequently heated, causing the remnants to become more
spherical with time \citep{Moore1996}.

Is environment the key factor in determining why two similar galaxies
may or may not have a bar or is the existence of a bar related to the
initial conditions of galaxy formation?  The stability (or
instability) of disks to bar formation may also depend on the baryon
fraction, and in particular the mass of stars and gas in the
disk. This varies across the Hubble sequence and depends strongly on
halo mass \citep{vandenBosch2000, Courteau2003, McGaugh2005}. This all
suggests that in addition to local density, morphology and halo mass
are also important principal parameters to investigate.  Another key
observational result is the fact that the bar fraction is not changing
significantly with redshift (see \citealt{Elmegreen2004,
  Marinova2007}, but also \citealt{Sheth2008} for a different result),
however, most disk galaxies are not within dense environments so it
would be difficult to disentangle the effects of environment,
especially at higher redshifts.
 
Several studies have shown that there is no evidence for a dependence
of bar frequency on galaxy environment \citep{VanDenBergh2002}, the
same is true even if galaxies of different morphological type are
considered independently \citep{Aguerri2009}. \cite{Li2009} came to
the same conclusion analyzing the clustering properties of barred and
unbarred galaxies of similar stellar mass and finding it
indistinguishable over all the scales probed (from $\sim$20 kpc to 30
Mpc). More recently, the Coma cluster was studied by
\cite{MendezAbreu2010}: they find that the bar fraction does not vary
significantly even when going from the center to the cluster
outskirts. However, the Coma cluster is such an extreme environment
that most of its apparent spiral galaxy population may be field
galaxies in projection.

In the light of these observational results and motivation from
numerical simulation studies, we aim at measuring the bar fraction (as
number of barred discs over the total number of discs) as a function
of environment and disc morphology, at $z\sim0$ in two carefully
selected samples representative of a low-density environment (the
isolated galaxies from the AMIGA sample) and of a moderately dense
environment (galaxies in the Virgo cluster).

To achieve this goal it is important to use homogeneous
classifications since, as we have shown in Giordano et al., (2010)
(paperI hereafter), the bar fraction is very stable against sample
selection but that some (possibly spurious) differences can arise if
the comparison is based on samples classified using different methods
(for example visual classification versus automated profile
fitting). In particular, the way the disc population is identified and
isolated plays a crucial role, since, if no detailed morphological
information is available, discs can easily be miscounted (for example
applying only color and/or magnitude cuts).

In order to address this, we use data from the UKIDSS Large Area
Survey \citep{Lawrence2007} and from SDSS DR7 \citep{Abazajian2009},
with the great advantage of combining optical {\texttt rgb} images
with near-infrared (H-band) imaging with excellent resolution for
local universe studies, that allow us to visually inspect the images
to provide detailed morphological classifications.

\begin{figure}     
\centering
\includegraphics[width=0.45\textwidth]{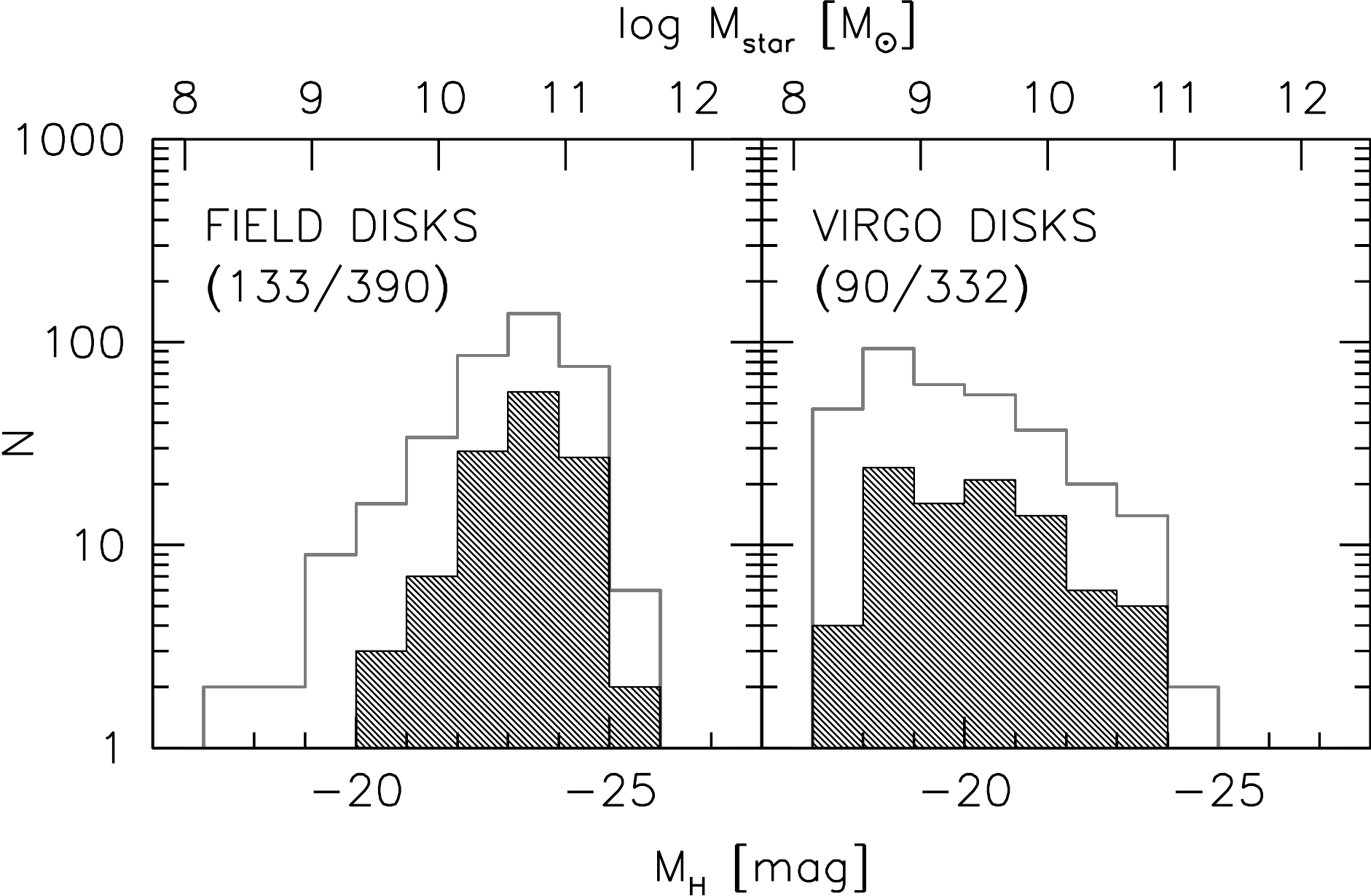}
\caption{H-band luminosity distribution of the \field\ and \cluster\
  moderately inclined (axis ratio $> 0.4$), morphologically selected
  discs (with Hubble type ranging from S0 to Sm). Shaded histograms
  represent the barred discs.}
\label{FIG:1}
\end{figure}

The outline of the paper is the following: in section \S
\ref{sec.data} we present the data that we are using, their
classification and the selection of the samples based on local density
estimation. The results about the bar fraction in the different cases
are presented in section \S \ref{sec.results} and discussed in section
\S \ref{sec.discussion}.

\section{Data}
\label{sec.data}

\subsection{\cluster\ sample}

In PaperI we presented a thorough study of the barred galaxies in the
Virgo Cluster from which we adopt all the classified galaxies with a
measured H-band magnitude from 2MASS. The \cluster\ disk sample is
composed of moderately inclined (axis ratio larger than 0.4) members
with UKIDSS near-IR imaging of Hubble type between S0 and Sm, spanning
a H-band magnitude (stellar mass) range of -17 to -25 mag ($10^8$ to
$10^{12}$ M$_{\odot}$).  In the following analysis, we use the H-band
magnitudes from Paper~I to compute stellar masses assuming a flat
$(B-H)$ color with a $\Upsilon_{H,*}=1$.  The local galaxy density for
members is determined via the $\rho_5$ proxy \citep{Baldry2006}, using
the positions and magnitudes from the Virgo Cluster Catalog
\citep{Binggeli1984_P1}.

\begin{figure}      
\centering
\includegraphics[width=0.45\textwidth]{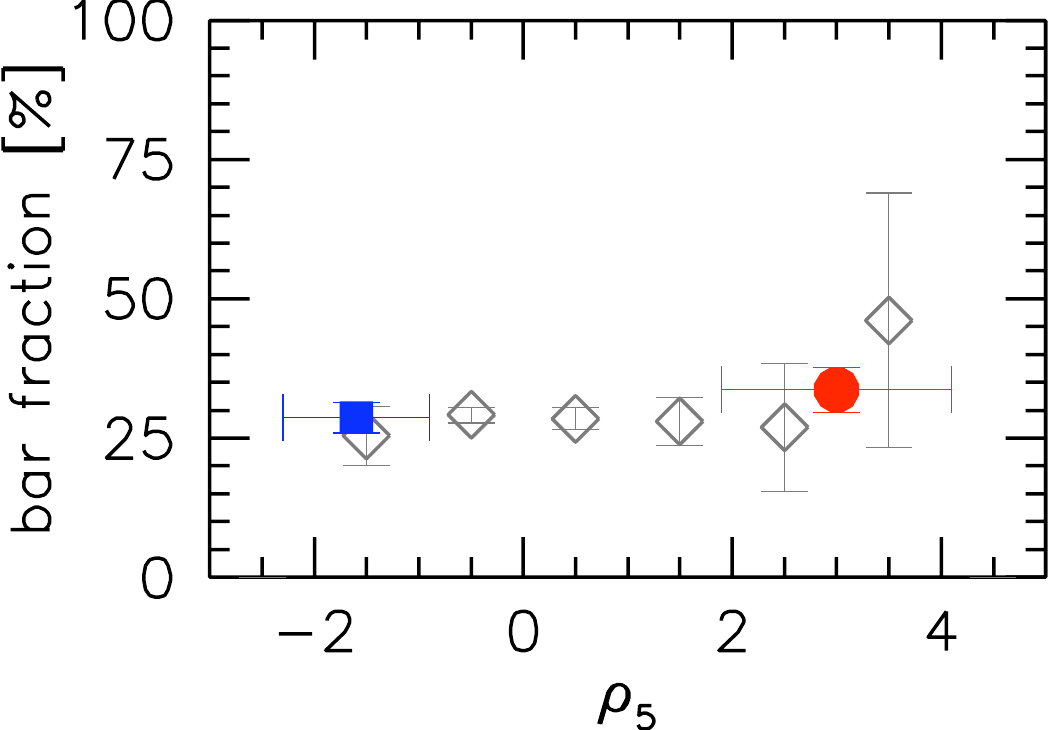}
\caption{Bar fraction as a function of local density measured via the
  $\rho_5$ proxy of the \sdssnair\ (grey diamonds), \field\ (blue
  square) and \cluster\ (red circle) samples, with \field\ and and
  \cluster\ samples bracketing the extremes in $\rho_5$ of the
  \sdssnair\ density distribution. Averaged over all morphological
  types, the bar fraction is constant against local density.}
\label{FIG:2}
\end{figure}

\subsection{\field\ sample}

To provide a robust comparison to the \cluster\ sample, we select a
true \field\ sample using the AMIGA (Analysis of the interstellar
Medium of Isolated GAlaxies) project
\citep{Verdes-Montenegro2005_A01}. The AMIGA catalogue is based on the
KIG catalog \cite{Karachentseva1973} of isolated galaxies ($z \lesssim
0.1$).  The KIG catalog is composed of 1050 galaxies with apparent
blue magnitudes brighter than 15.7 mag; these isolated galaxies are
selected to have no neighbor of comparable size within twenty galactic
diameters.  The KIG catalog has been used by multiple studies to
investigate the effects of under-dense environment on galaxy
properties \citep{Adams1980, Haynes1980} and the AMIGA project
quantified the isolation of KIG galaxies identifying their sample of
791 genuinely isolated galaxies \cite{Verley2007_A05}. The AMIGA
project has also compiled multiwavelength coverage of this
statistically significant sample of the most isolated galaxies in the
local universe and the dataset includes optical photometry and
morphologies for a redshift-complete subsample of 956 galaxies
\citep{Sulentic2006_A02}, and these data are publicly released under
a VO interface at \texttt{http://amiga.iaa.es/}.

By cross-matching the AMIGA catalogue with the 2MASS and SDSS
databases, we identify 563 galaxies with both H-band and {\texttt rgb}
images. As in our \cluster\ sample, we select only moderately inclined
(axis ratio $>0.4$) disk galaxies with Hubble type of $S0$ to $Sm$.
The stellar masses of the resulting 390 \field\ disk galaxies are
determined using the total H-band magnitudes from 2MASS and assuming
the same flat $(B-H)$ color and $\Upsilon_{H,*}=1$ as in the \cluster\
sample.  The local density for each galaxy is determined by their 5th
nearest neighbors ($\rho_5$, see \citealt{Baldry2006}) as defined by
the AMIGA catalogue of neighbors, constructed down to a magnitude
$m\sim$17.5, lying within 0.5 Mpc around the KIG galaxies
\citep{Verley2007_A04}.

\subsection{Identifying Barred Disks}
\label{subsec:bars}

The UKIDSS Large Area Survey is an ongoing survey to image 4000
deg$^2$ at high Galactic latitudes in the YJHK filters to a depth in H
of 18.8 mag; it has a spatial resolution of 0.4$^{\prime
  \prime}$/pixel (like SDSS data) and an average seeing of
0.8$^{\prime \prime}$. Our galaxies span a redshift out to $z
\sim$0.03 where the UKIDSS imaging has a physical resolution of
$\sim$400 pc/pixel.  As outlined in PaperI, we use the H-band imaging
to visually classify the disk galaxies into one of three categories:
``barred'', ``non-barred'', or ``uncertain''. All the galaxies in the
\cluster\ sample have H-band imaging from UKIDSS that is also
available for 172 galaxies in the \field\ sample, for the rest we must
rely on SDSS $z^{\prime}$ and \texttt{rgb} imaging; however, for all
the galaxies with both H-band and $z'$ imaging, we find that our bar
classifications are essentially identical.

Both \cluster\ and \field\ catalogues, comprehensive of \texttt{rgb}
and H-band thumbnails are available
online\footnote{http://www.itp.uzh.ch/~giordano/mwpbg-v2.html}.

\begin{figure*}     
\centering
\includegraphics[width=0.84\textwidth]{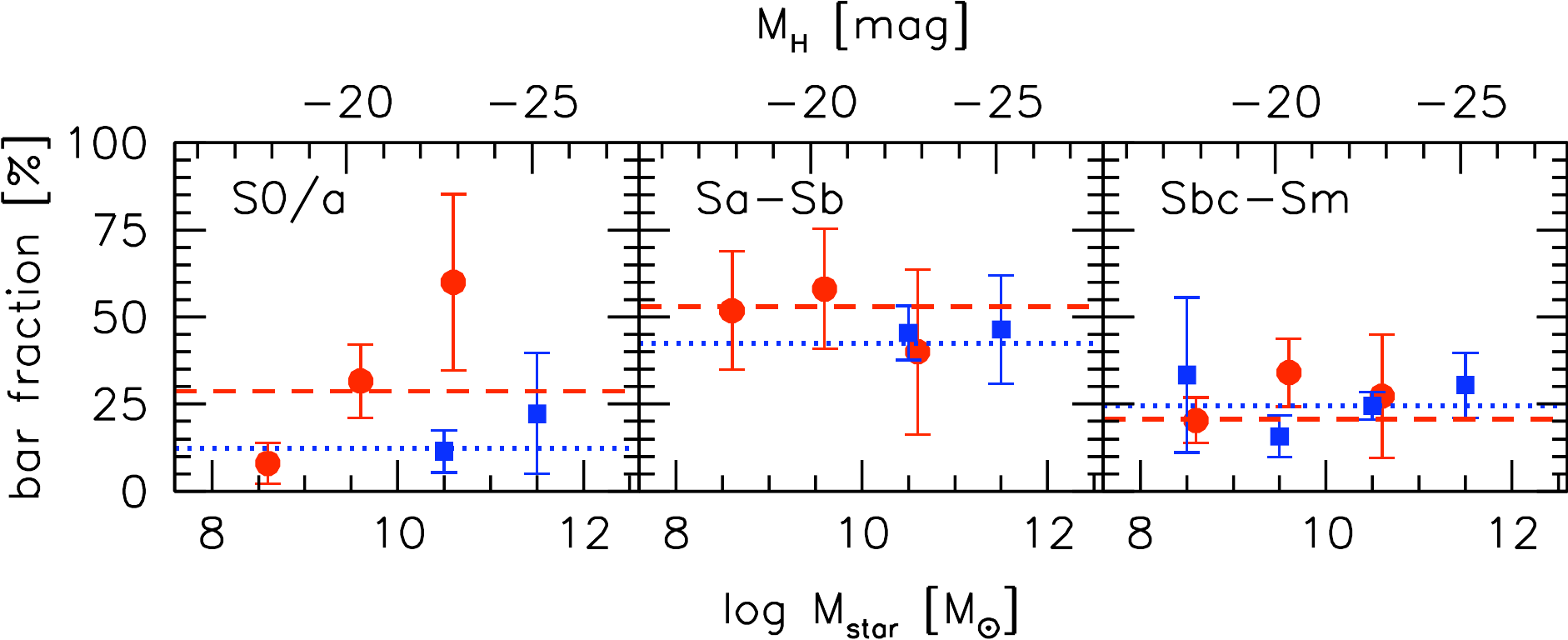}
\caption{Bar fraction vs. stellar mass distribution of the \field\ and
  \cluster\ morphologically selected disks. The red circles represent
  the \cluster\ sample, the blue squares the \field\ sample. In each
  panel, the average value over the morphological type is represented
  by the dashed/dotted line, for the \cluster\ (red) and the \field\
  (blue) respectively. The $Sa-Sb$ disks have the highest bar
  fraction, regardless of environment, suggesting that baryon fraction
  is the main driver for bar instabilities.}
 \label{FIG:3}
\end{figure*}

\vskip 1.2cm

\section{Results and Discussion}
\label{sec.results}

\subsection{Barred Fraction vs. Environment}

In Figure \ref{FIG:1} we show the H-band luminosity distribution for
moderately inclined disk galaxies in our \field\ and \cluster\
samples. The barred fraction averaged over the disk population in the
\field\ is $\sim$34\% (133/390) and in \cluster\ is $\sim$28\%
(90/332).  We find that even with bar classifications based on high
resolution near-IR imaging, the barred fraction does not vary with
environment when considering the disk population as a whole.  Although
our results are consistent with earlier work based on optical imaging
(see for example \citealt{VanDenBergh2002,Li2009}), the constancy of
the barred fraction conflicts with expectations from current galaxy
formation models, since strong interactions can trigger bar
instabilities \citep{Berentzen2004}.

To further test the robustness of our result, we incorporate the
\cite{Nair2010} optically-selected sample of about 14,000 galaxies
from the SDSS DR4; this sample includes nearly all
spectroscopically-targeted galaxies in the redshift range $0.01 < z <
0.1$ to an apparent extinction-corrected limit of $g<16$ mag. In
addition to visually classified Hubble types, the \sdssnair\ catalogue
contains the existence of bars, stellar masses computed according to
\cite{Kauffmann2003a}, and an estimate of the local density computed
using the 5th nearest neighbors ($\rho_5$) according to
\cite{Baldry2006}.  To ensure that we can compare the (optical)
\sdssnair\ sample directly to our (near-IR) results, we apply the same
disk selection criteria as in our \field\ and \cluster\ samples and
consider only galaxies at $z \leq 0.03$.
 
Figure \ref{FIG:2} shows the barred fraction in our \field\ (blue) and
\cluster\ (red) populations as well as the \sdssnair\ sample (grey) as
a function of local galaxy density ($\rho_5$); note how effectively
our \field\ and \cluster\ samples bracket the extremes in $\rho_5$.
We find the barred disk fraction is surprisingly resilient
($\sim30-35$\%): the barred fraction does not vary as the environment
changes from isolated field galaxies to cluster cores.

\subsection{Barred Fraction vs. Disk Morphology}

To investigate how the barred fraction varies with disk morphology, we
divide both our \field\ and \cluster\ samples into three classes: 1)
lenticulars (featureless discs, corresponding to Hubble-types
$S0-S0a$); 2) early-type spirals (bulge-dominated discs, corresponding
to Hubble-types $Sa-Sb$); and 3) late-type spirals (disc-dominated or
bulge-less discs, corresponding to Hubble-types $Sbc-Sm$).  In Figure
\ref{FIG:3} we show the barred fraction as a function of stellar mass
for the three disk classes where the blue squares represent the
\field\ and the red circles the \cluster\ members.  Each panel also
includes the average barred fraction for the three disk classes in the
\field\ (dotted line) and in \cluster\ (dashed line).  The differences
in the relative number of galaxies in each disk class is due to the
morphology-density relation, i.e. the fraction of lenticulars in the
\field\ ($<10$\%) is lower than in \cluster\ ($\sim30$\%; PaperI).

We find that the barred fraction for early-type spirals is
systematically higher than in late-type spirals {\it regardless of
  environment}: $45-50$\% for $Sa-Sb$ vs. $<25$\% for $Sbc-Sm$.  The
barred fraction in lenticulars ($S0-S0a$) is also lower in both
environments.

\section{Summary}
\label{sec.discussion}

We present the first comprehensive study of barred disks as a function
of environment that uses NIR and \texttt{rgb} imaging to resolve bars;
the advantage of using near-infrared imaging from UKIDSS is that bar
classifications are less affected by dust and bright star-forming
regions.  We expand on our study of bars in \cluster\ by building a
\field\ sample using the KIG catalog of isolated galaxies.  Our
\field\ and \cluster\ disk populations are at $z<0.03$, span a range
in stellar mass from $\sim$10$^8$ to $\sim$10$^{12}$ M$_{\odot}$ and
Hubble type ($S0-Sm$), encompass a wide range in local densities and
are analyzed in exactly the same manner.

We find that the barred disk fraction is surprisingly constant at
$30-35$\% in both the \field\ and \cluster\ samples, i.e. the barred
fraction for the disk population as a whole does not depend on
environment.  We test the robustness of our result by analyzing the
NA10 optically-selected sample of nearby galaxies in the same manner,
and we again find a constant barred fraction across the full range of
local galaxy density.

This implies that disks become barred prior to the late time assembly
of galaxy clusters, which is consistent with observational evidence
that the bar fraction does not evolve strongly with redshift.

The barred fraction is highest for early-type spirals ($Sa-Sb$)
regardless of environment: these galaxies are nearly twice as likely
to be barred as late-type spirals ($Sbc-Sm$).  If a late type spiral
forms a bar, then it may also form a pseudo-bulge via a buckling
instability and its morphological class will change.  Indeed, the
consensus is forming that our own Galaxy has evolved across the Hubble
sequence in this fashion \citep{Oski2010}.  If this is a common
phenomenon, as numerical simulations indicate \citep{Debattista2006},
then we naturally expect the bar fraction to be higher in early type
spirals, that have a higher baryon fraction.  This implies that a
significant fraction of the bulges of early type galaxies are
pseudo-bulges.

The morphology-density relation \citep{Dressler1980} can be explained
by the notion that the cluster environment is creating S0's from the
infalling disc population. Indeed, \cite{Graham2008} find that the
bulge to disc ratios of S0's is similar to that of early type
galaxies.  One might therefore expect the bar fraction to be the same
in S0s and early type spirals, however averaged over the entire
population it is significantly lower (25\% versus 50\% respectively).
We note that the bar fraction in S0's and early type discs with
stellar masses above $10^{10}M_\odot$ is similar ($\sim$ 50\%), but
this drops to less than 10\% in the least massive S0's. This supports
a harassment scenario for the formation of the S0
population. Gravitational encounters between galaxies and with the
global cluster potential thicken the disks of massive early type
spirals by an amount that is sufficient to suppress spiral patterns
(Moore et al 1999). For lower mass disks, the heating is more
effective and will eventually erase the signatures of a preexisting
bar. Numerical simulations also indicate that infalling late type
disks will undergo an environmentally driven bar instability, however
this phase is short lived with the bar experiencing subsequent heating
until it becomes a dE/dSph.


\clearpage

\end{document}